\documentclass[lettersize,journal]{IEEEtran}
\usepackage{cite}
\usepackage{amsmath,amsfonts}
\usepackage{graphicx}

\usepackage{textcomp}
\usepackage{xcolor}
\usepackage{CJKutf8}
\usepackage{tabularray}
\usepackage{subfigure}
\usepackage{caption}
\usepackage{multirow}
\usepackage{bm}
\usepackage{amsmath,amsfonts}
\usepackage{algorithm}
\usepackage{array}
\usepackage[caption=false,font=normalsize,labelfont=sf,textfont=sf]{subfig}
\usepackage{textcomp}
\usepackage{stfloats}
\usepackage{url}
\usepackage{verbatim}
\usepackage{graphicx}
\usepackage{cite}
\usepackage{amsmath,amsfonts}
\usepackage{array}
\usepackage[caption=false,font=normalsize,labelfont=sf,textfont=sf]{subfig}
\usepackage{textcomp}
\usepackage{stfloats}
\usepackage{url}
\usepackage{verbatim}
\usepackage{graphicx}
\def\BibTeX{{\rm B\kern-.05em{\sc i\kern-.025em b}\kern-.08em
    T\kern-.1667em\lower.7ex\hbox{E}\kern-.125emX}}
\usepackage{balance}
\usepackage{amsmath}
\usepackage{makecell}
\usepackage{array}
\newcolumntype{C}{>{\centering\arraybackslash}X}
\usepackage{epstopdf}

\usepackage{bm}
\usepackage{changes}
\usepackage{textcomp}
\usepackage{xcolor}
\usepackage{tabularx,booktabs}
\usepackage[flushleft]{threeparttable}
\usepackage{booktabs,ragged2e}
\usepackage{multirow}
\usepackage{subcaption}
\usepackage{bm}

\usepackage{amsthm}

\allowdisplaybreaks[4]

\begin{document}

\begin{CJK}{UTF8}{gbsn}
\title{Privacy-Preserving Semantic Communication over Wiretap Channels with Learnable Differential Privacy}


\author{Weixuan Chen,~\IEEEmembership{Graduate Student Member, IEEE}, Qianqian Yang,~\IEEEmembership{Member, IEEE}, Shuo Shao,~\IEEEmembership{Member, IEEE}, Shunpu Tang,~\IEEEmembership{Student Member, IEEE},
Zhiguo Shi,~\IEEEmembership{Fellow, IEEE}, and Shui Yu,~\IEEEmembership{Fellow, IEEE}

\thanks{

This paper was presented partially in \textit{Proc. 34th IEEE MLSP, London, UK, Sep. 2024}\cite{chen2024enhancing}.

W. Chen, Q. Yang$^{\dag}$, S. Tang, and Z. Shi are with the College of Information Science and Electronic Engineering, Zhejiang University, Hangzhou,
China. (e-mails: \{weixuanchen, qianqianyang20$^{\dag}$, tangshunpu, shizg\}@zju.edu.cn).

S. Shao is with the Department of System Science, University of Shanghai for Science and Technology, Shanghai, China. (e-mail: shuoshao@usst.edu.cn).

S. Yu is with the Faculty of Engineering and Information Technology, University of Technology Sydney, Australia. (e-mail: Shui.Yu@uts.edu.au).

This work is partly supported by the NSFC under grant No. 62293481, No. 62571487, No. 62201505, by the National Key R\&D Program of China under Grant 2024YFE0200802, and by the Zhejiang Provincial Natural Science Foundation of China under Grant No. LZ25F010001.  (Corresponding author: Qianqian Yang.)



}

}

\maketitle

\begin{abstract}

%
%

%
%
%

While semantic communication (SemCom) improves transmission efficiency by focusing on task-relevant information, it also raises critical privacy concerns. 
Many existing secure SemCom approaches rely on restrictive or impractical assumptions, such as favorable channel conditions for the legitimate user or prior knowledge of the eavesdropper's model.
To address these limitations, this paper proposes a novel secure SemCom framework for image transmission over wiretap channels, leveraging differential privacy (DP) to provide approximate privacy guarantees.
Specifically, our approach first extracts disentangled semantic representations from source images using generative adversarial network (GAN) inversion method, and then selectively perturbs private semantic representations with approximate DP noise.
Distinct from conventional DP-based protection methods, we introduce DP noise with learnable pattern, instead of traditional white Gaussian or Laplace noise, achieved through adversarial training of neural networks (NNs).
This design mitigates the inherent non-invertibility of DP while effectively protecting private information.
Moreover, it enables explicitly controllable security levels by adjusting the privacy budget according to specific security requirements, which is not achieved in most existing secure SemCom approaches.
Experimental results demonstrate that, compared with the previous DP-based method and direct transmission, the proposed method significantly degrades the reconstruction quality for the eavesdropper, while introducing only slight degradation in task performance.
Under comparable security levels, our approach achieves an LPIPS advantage of 0.06-0.29 and an FPPSR advantage of 0.10-0.86 for the legitimate user compared with the previous DP-based method.

%
%

\end{abstract}


\section{Introduction}

%

\subsection{Backgrounds}

Semantic communication (SemCom) \cite{zhang2024semanticsurvey} has recently emerged as a promising paradigm for future 6G networks. 
Compared with conventional digital communication systems that aim to ensure bit-level accuracy, SemCom focuses on transmitting only task-relevant information.
%
Building upon this paradigm, recent studies have demonstrated that SemCom can significantly enhance communication efficiency across various data modalities, including text \cite{han2022semantictext,peng2024robusttext}, speech \cite{han2022semantic,chen2024lowcomspeech}, images \cite{chen2023deep,tang2024contrastive}, and video \cite{wang2022wireless,guo2025videoqasc}.
Benefiting from these advances, SemCom is expected to enable future application scenarios such as digital twins, immersive communication, and embodied or agentic AI.

However, the SemCom approach typically eliminates traditional channel coding, which introduces redundancy that can provide a certain degree of protection for the transmitted information. 
Moreover, SemCom prioritizes transmitting semantically significant information while discarding as much redundancy as possible. 
As a result, private information, which often carries high semantic importance, becomes more
vulnerable to exposure by unauthorized users.
Many studies have explored secure SemCom over wiretap channels \cite{shen2023secure,li2024secure,wang2024tifs,jincheng2025privacy,rong2025tifs}, which has become an important research direction and has received increasing attention in recent years.

\subsection{Related Works and Motivation}


Researchers have explored a wide range of solutions to protect semantic information from eavesdropping. 
These solutions can generally be categorized into three main approaches: \textit{adversarial training-based methods}, \textit{encryption-based methods}, and \textit{physical layer-based methods}. 
Next, we briefly review each of these approaches in the following.

\subsubsection{Adversarial training-based methods}

%
%

%

Marchioro \textit{et al.} \cite{marchioro2020adversarial} proposed an adversarially trained joint source-channel coding (JSCC) framework that models the transmitter-receiver pair and the eavesdropper as players in a minimax game, aiming to confuse the eavesdropper's classifier while preserving task performance.
Following a similar adversarial design, 
Shi \textit{et al.} \cite{shi2025secure} introduced a controlled minimax objective that constrains the eavesdropper's performance within a specified threshold.
Other studies have explored alternative JSCC variants to achieve robust adversarial learning.
Erdemir \textit{et al.} \cite{erdemir2022privacy} proposed a VAE-based JSCC scheme trained end-to-end to capture the privacy-utility trade-off (PUT), 
while Zhang \textit{et al.} \cite{zhang2023wireless} developed a residual-convolutional autoencoder-based JSCC framework with a SecureMSE loss, optimized to drive the eavesdropper's reconstruction toward meaningless outputs.
However, these approaches generally rely on the impractical assumption that the legitimate user has knowledge of the eavesdropper's model, and most lack mechanisms to explicitly control the level of system security.

\subsubsection{Encryption-based methods} 

%

%
%

%

Some efforts have explored deep learning-based encryption.
Luo \textit{et al.} \cite{luo2023encrypted} proposed an encrypted SemCom system (ESCS), where the semantic message and a symmetric key are jointly processed by a neural network (NN)-based encryptor-decryptor trained adversarially to ensure confidentiality.
Building upon this, Qin \textit{et al.} \cite{qin2023securing} proposed a key generation method based on the randomness of BLEU scores in machine translation and introduced a subcarrier-level obfuscation mechanism using these semantic keys. 
Expanding further on cryptographic formulations,
Tung \textit{et al.} \cite{tung2023deep} proposed a framework that integrates a public-key encryption scheme based on the learning with errors (LWE) problem into a JSCC autoencoder, ensuring ciphertext indistinguishability under chosen-plaintext attacks.
Moreover, Meng \textit{et al.} \cite{meng2025secure} investigated the use of homomorphic encryption for secure SemCom.
They demonstrated that semantic information can be preserved within encrypted data and designed a privacy-preserving JSCC model that supports homomorphic operations, achieving comparable task performance on both plaintext and ciphertext. 
However, these approaches may suffer from high computational overhead caused by sophisticated encryption schemes or from complex key management.

%
%
%

\subsubsection{Physical layer-based methods}

%
Researchers have explored introducing artificial noise into the transmitted signals in the physical layer.
Chen \textit{et al.} \cite{chen2024nearly} investigated power control for artificial noise to achieve specific security levels.
In their scheme, semantic information and random symbols are superposed onto a double-layered constellation. 
By adjusting the power allocation in superposition coding to control the symbol error probabilities (SEPs) of both users, they achieved nearly zero information leakage to the eavesdropper while maintaining reliable task performance.
Building on this concept, Chen \textit{et al.} \cite{chen2025knowledge} further developed a coding-enhanced jamming scheme that enables the legitimate receiver to cancel the artificial noise using shared private knowledge, while the eavesdropper cannot.
Specifically, they generated private codebooks based on the shared knowledge and combined semantic information with jamming signals derived from these codebooks through superposition coding. 
Another approach explores the use of diffusion models.
%
He \textit{et al.} \cite{he2025diffusionenabledsecure} designed and injected artificial Gaussian noise or adversarial perturbations into the transmitted signals, and employed denoising diffusion probabilistic models (DDPM) to suppress both adversarial and channel noise for the legitimate user.
However, these approaches may rely on the legitimate user's channel advantage or lack theoretically provable privacy guarantees.

From the above discussion, it can be observed that existing approaches still face fundamental limitations.
Most of them achieve security in specific settings or through strong assumptions, such as channel advantage, model knowledge, or pre-shared keys, which are often impractical in real systems.
Moreover, current designs lack a unified mechanism to flexibly control the level of security while maintaining communication efficiency. 
To bridge this gap, we are motivated to design a SemCom system that provides explicitly controllable security in the challenging comparable-SNR wiretap channel scenario \cite{wyner1975wire}, where the eavesdropper and the legitimate receiver experience comparable channel SNRs, while eliminating the need for key exchange.

To achieve this, we are inspired to introduce differential privacy (DP) \cite{dwork2006differential} into SemCom, as DP is a well-established framework that provides quantifiable privacy guarantees by injecting controlled noise to obscure sensitive information while preserving data utility.
While previous studies have explored applying DP to protect source data, such as visual or semantic representations \cite{fan2018image, fan2019practical, li2021differentially,xue2021dp,wen2022identitydp}, directly applying DP to SemCom over wiretap channels remains challenging. 
This is because the inherent non-invertibility of DP, together with the presence of channel noise, often leads to severe degradation in reconstruction quality at the legitimate receiver, thereby compromising communication reliability.

\subsection{Contributions}

In this paper, we propose a novel secure SemCom system that leverages the DP technique to enhance security when the legitimate user and the eavesdropper experience comparable channel conditions.
To efficiently extract semantic information and distinguish privacy-sensitive content, we first employ generative adversarial network (GAN) inversion to obtain disentangled semantic representations. 
The sensitive semantic representations are selectively determined using pre-defined indices and perturbed with learnable DP noise, while the non-sensitive parts remain unchanged to preserve semantic fidelity. 
This pre-definition is performed once and shared between the transmitter and the legitimate receiver, enabling fine-grained protection without key exchange.

%

To overcome the non-invertibility problem of traditional DP mechanisms, we further propose novel NN-based DP protection and deprotection modules.
The protection module generates learnable DP noise patterns through adversarial training, ensuring that the added perturbation statistically resembles genuine DP noise for privacy preservation. 
Meanwhile, the deprotection module at the legitimate receiver learns to recognize and mitigate these structured noise patterns, effectively restoring the protected semantics.
In addition, the intensity of the injected DP noise can be adaptively adjusted according to different privacy budgets, thereby enabling an explicit trade-off between privacy protection and reconstruction quality.

The main contributions of this paper are summarized as follows:


%
%
\begin{itemize}

\item We investigate a novel secure SemCom framework that leverages DP to ensure communication security in a challenging comparable-SNR wiretap channel scenario. 
By extracting disentangled semantic representations through GAN inversion and selectively perturbing privacy-sensitive features according to pre-shared indices, the proposed system achieves fine-grained semantic protection while eliminating the need for key exchange and maintaining high reconstruction fidelity for the legitimate user.

\item We propose NN-based DP protection and deprotection modules to address the non-invertibility and performance degradation problems of conventional DP mechanisms. 
Through adversarial optimization, the protection module learns to generate DP noise with learnable pattern that is statistically similar to standard DP noise, while the deprotection module effectively removes it at the receiver. 
Moreover, by tuning the privacy budget, our method enables explicitly controllable security levels, which most existing methods lack, enabling flexible adaptation to various privacy requirements and communication scenarios.

\item We conduct extensive experiments to demonstrate that, compared with the previous DP-based method~\cite{li2021differentially} and SemCom without any security mechanism, the proposed framework achieves higher system security while maintaining high-fidelity reconstruction for the legitimate user and robust adaptability under varying channel conditions. In particular, under comparable security levels, our approach achieves an LPIPS improvement of 0.06-0.29 and a face privacy protection success rate (FPPSR) improvement of 0.10-0.86 for the legitimate user compared with the previous DP-based method.


%
%
%
%
%
\end{itemize}

\section{System Model}

\begin{figure*}[t]
\begin{center}
\centerline{\includegraphics[width=0.85\linewidth]{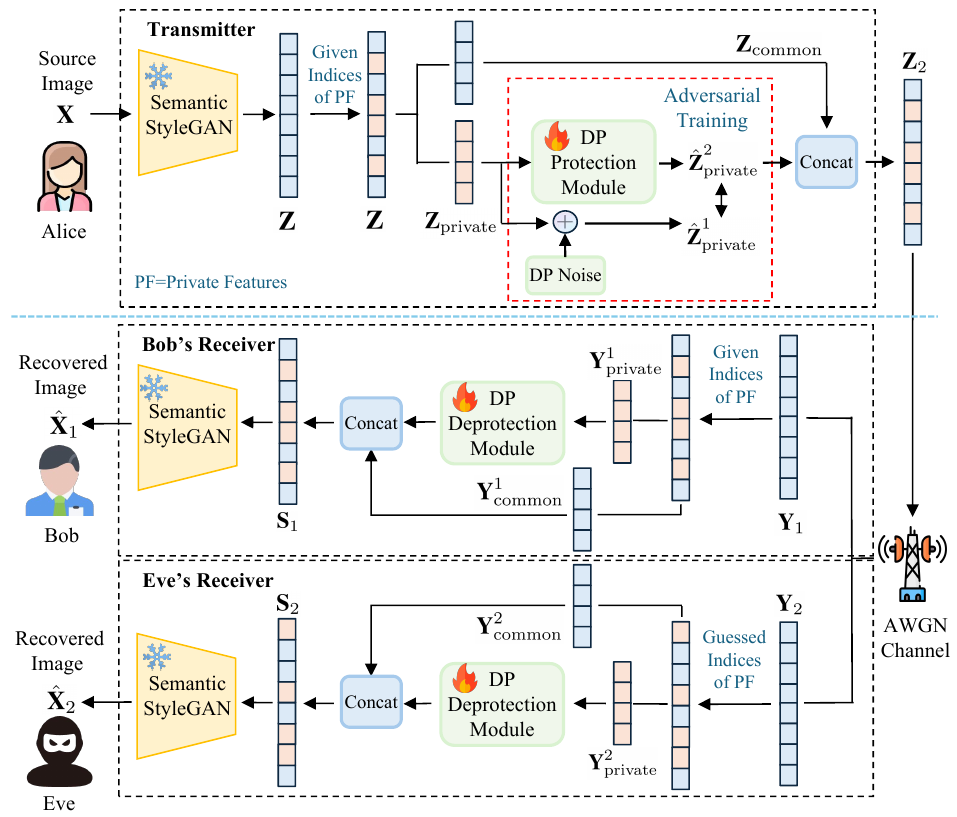}}
\caption{The framework of our proposed secure SemCom system.}
\label{fig.1}
\end{center}
\vskip -0.3in
\end{figure*}

We consider a SemCom system designed for secure image transmission over wiretap channels, as illustrated in Fig.~\ref{fig.1}. 
The system comprises three entities: a transmitter (Alice), a legitimate receiver (Bob), and an eavesdropper (Eve). 
In this system, Alice seeks to transmit a source image, denoted by $\textbf{X}$, to Bob over a noisy wireless channel. 
To preserve privacy, Alice protects the sensitive portion of the image, denoted as $\textbf{X}_{\rm private}$, while the non-sensitive part, $\textbf{X}_{\rm common}$, remains unprotected.
Meanwhile, an eavesdropper, Eve, passively intercepts the transmission through a channel of comparable quality, attempting to reconstruct or infer sensitive identity-related information from the intercepted signal.
%
%
%
%
%



Alice utilizes a transmitter to encode and protect the source image $\textbf{X}$, extracting the semantic representation (also referred to as latent codes) to be transmitted, denoted by $\textbf{Z}_2$:
\begin{equation}
    {\textbf{Z}}_{2} = f_{\mathrm{Alice}}\left( \textbf{X} \right),
\end{equation}
where $f_{\mathrm{Alice}}$ denotes the semantic encoder. 
The semantic representation $\textbf{Z}_2$ is normalized to satisfy the average power constraint $P$, and then mapped into a complex vector $\tilde{\textbf{Z}}_2$ by pairing values into complex symbols for transmission over an AWGN channel. 
%
%
%
%
Bob and Eve receive noisy semantic representations, given by
\begin{equation}
    \textbf{Y}_{1} = \tilde{\textbf{Z}}_2 + \textbf{n}_1,
\end{equation}
\begin{equation}
    \textbf{Y}_{2} = \tilde{\textbf{Z}}_2 + \textbf{n}_2,
\end{equation}
where $\textbf{n}_{1/2} \sim \mathcal{N}(0, \sigma^{2}_{1/2})$ denotes Gaussian noise with zero mean and variance $\sigma^{2}_{1/2}$. Here, $\textbf{Y}_{1}$ and $\textbf{Y}_{2}$ denote the received semantic representations at Bob and Eve, respectively. 
The channel SNR between Alice and the legitimate user/eavesdropper is given by
\begin{equation}
    \mathrm{SNR}_{\rm leg/\rm eve} = 10 \log_{10}\left(\frac{P}{\sigma^{2}_{1/2}}\right) \, (\mathrm{dB}).
\end{equation}
In this paper, we assume that Eve experiences channel conditions comparable to those of Bob.

Both Bob and Eve attempt to reconstruct the source image $\textbf{X}$ as accurately as possible. The received signals $\textbf{Y}_{1}$ and $\textbf{Y}_{2}$ are first converted into real-valued vectors by separating the real and imaginary parts of each complex symbol. 
Bob decodes $\textbf{Y}_1$ to produce the recovered image, denoted by $\hat{\textbf{X}}_1$:
\begin{equation}
    \hat{\textbf{X}}_1 = f_{\mathrm{Bob}}\left( \textbf{Y}_1 \right),
\end{equation}
where $f_{\mathrm{Bob}}$ denotes Bob's decoder. Similarly, Eve decodes $\textbf{Y}_2$ into the recovered image $\hat{\textbf{X}}_2$:
\begin{equation}
    \hat{\textbf{X}}_2 = f_{\mathrm{Eve}}\left( \textbf{Y}_2 \right),
\end{equation}
where $f_{\mathrm{Eve}}$ denotes Eve's decoder.

This paper aims to develop a SemCom framework that fulfills two key objectives: (1) enabling Bob to accurately recover the original image, and (2) restricting the amount of sensitive semantic information that Eve can infer or reconstruct from the intercepted signals.
To assess the effectiveness of the proposed system, we employ both perceptual quality and privacy protection metrics. 
For image reconstruction quality, 
we adopt the learned perceptual image patch similarity (LPIPS) metric \cite{zhang2018unreasonable}, which measures the perceptual similarity between two images in the feature space of an image classification network.
As a perceptual metric, LPIPS provides a reliable approximation of human visual perception. Specifically, we adopt the AlexNet-based LPIPS \cite{krizhevsky2012imagenet}.

To assess privacy protection, we adopt the FPPSR. Specifically, we employ the ArcFace recognition system \cite{deng2019arcface}, which outputs a confidence score indicating whether two facial images belong to the same individual. The FPPSR is defined as the percentage of reconstructed faces identified as different from the original. Following empirical thresholds, a face is considered different if the ArcFace score falls below 0.31 \cite{xue2021dp}.

\section{Proposed Method}

\subsection{System Overview}

In this section, we provide an overview of the considered SemCom system, as illustrated in Fig.~\ref{fig.1}, which includes the transmitter, and the receivers at Bob and Eve.

\subsubsection{Transmitter}

At the transmitter, Alice utilizes the inversion method of a pre-trained Semantic StyleGAN \cite{shi2022semanticstylegan} $f_{\mathrm{inv}}$ to transform the source facial image, denoted by $\textbf{X}$, into a disentangled semantic representation $\textbf{Z}$, i.e., $\textbf{Z} = f_{\mathrm{inv}}\left( \textbf{X} \right)$.
This semantic representation comprises multiple disentangled latent codes, each associated with a specific attribute (e.g., eyes, nose, or mouth).
%

%
%
We denote the latent codes in $\textbf{Z}$ that require protection as the private latent codes (also referred to as private features), $\textbf{Z}_{\rm private}$, while the remaining ones are referred to as the common latent codes, $\textbf{Z}_{\rm common}$, i.e.,
%
\begin{equation}
    \textbf{Z} = \left[\textbf{Z}_{\rm private}, \textbf{Z}_{\rm common}\right].
\end{equation}
Then we apply the proposed privacy-preserving mechanism to protect $\textbf{Z}_{\rm private}$ to protect against interception by Eve. 
Specifically, we utilize an NN-based DP protection module to transform the private latent codes $\textbf{Z}_{\rm private}$ into the protected latent codes $\hat{\textbf{Z}}_{\rm private}^{2}$, i.e., 
%
\begin{equation}
    \hat{\textbf{Z}}_{\rm private}^{2} = f_{\mathrm{protection}}\left( \textbf{Z}_{\rm private};\bm{\theta}^{\mathrm{protection}} \right),
\end{equation}
where $f_{\mathrm{protection}}$ denotes the NN-based DP protection module, and $\bm{\theta}^{\mathrm{protection}}$ represents its learnable parameters.
%
The protected private latent codes $\hat{\textbf{Z}}_{\rm private}^{2}$ are then combined with the common latent codes $\textbf{Z}_{\rm common}$ to form the semantic representation to be transmitted, 
denoted as $\textbf{Z}_2$, i.e.,
\begin{equation}
    \textbf{Z}_2 = \left[\hat{\textbf{Z}}_{\rm private}^{2}, \textbf{Z}_{\rm common}\right].
\end{equation}

\subsubsection{Bob's Receiver}

Bob knows the true indices of the private latent codes. 
Based on this knowledge, the real vector $\textbf{Y}_1$ is divided into the private and common components, $\textbf{Y}_{\rm private}^1$ and $\textbf{Y}_{\rm common}^1$, respectively, following the same partitioning as at the transmitter.
Next, $\textbf{Y}_{\rm private}^1$ is passed through an NN-based DP deprotection module to mitigate the effect of the applied protection, yielding the refined private latent codes $\hat{\textbf{Y}}_{\rm private}^{1}$: 
\begin{equation}
    \hat{\textbf{Y}}_{\rm private}^{1} = f_{\mathrm{deprotection}}\left( \textbf{Y}_{\rm private}^1; \bm{\theta}_{1}^{\mathrm{deprotection}} \right),
\end{equation}
where $f_{\mathrm{deprotection}}$ denotes the NN-based DP deprotection module at the legitimate user, and $\bm{\theta}_{1}^{\mathrm{deprotection}}$ represents its learnable parameters.
The refined private latent codes $\hat{\textbf{Y}}_{\rm private}^{1}$ are then combined with the common latent codes $\textbf{Y}_{\rm common}^1$ to form the complete semantic representation $\textbf{S}_{1}$:
\begin{equation}
    \textbf{S}_1 = \left[\hat{\textbf{Y}}_{\rm private}^{1}, \textbf{Y}_{\rm common}^1\right].
\end{equation}
Finally, Bob inputs $\textbf{S}_{1}$ into a pre-trained Semantic StyleGAN generator $f_{\mathrm{gen}}$ to reconstruct the source image, i.e.,
\begin{equation}
    \hat{\textbf{X}}_{1} = f_{\mathrm{gen}}\left( \textbf{S}_{1} \right).
\end{equation}
where $\hat{\textbf{X}}_{1}$ denotes the reconstructed image. 

\subsubsection{Eve's Receiver}

We consider two possible scenarios for Eve. 
In the first scenario, Eve is unaware that the semantic representation $\tilde{\textbf{Z}}_2$ has been protected. 
As a result, Eve directly inputs $\textbf{Y}_2$ into the pre-trained Semantic StyleGAN generator to reconstruct the image $\hat{\textbf{X}}_2$, i.e., 
\begin{equation}
    \hat{\textbf{X}}_{2} = f_{\mathrm{gen}}\left( {\textbf{Y}_2} \right).
\end{equation}

In the second scenario, Eve is aware that the semantic representation $\tilde{\textbf{Z}}_2$ has been protected and has stolen the architecture of the DP deprotection module. 
However, Eve does not know the trained parameters of the deprotection module, nor does she know the indices of the protected latent codes.
In this case, based on the guessed indices of private latent codes, $\textbf{Y}_{2}$ is divided into the private and common components, $\textbf{Y}_{\rm private}^2$ and $\textbf{Y}_{\rm common}^2$.  
$\textbf{Y}_{\rm private}^2$ is then passed through Eve's NN-based DP deprotection module to mitigate the effect of protection:
\begin{equation}
    \hat{\textbf{Y}}_{\rm private}^{2} = g_{\mathrm{deprotection}}\left( {\textbf{Y}}_{\rm private}^2; \bm{\theta}_{2}^{\mathrm{deprotection}} \right),
\end{equation}
where $g_{\mathrm{deprotection}}$ represents Eve's DP deprotection module, and $\bm{\theta}_{2}^{\mathrm{deprotection}}$ denotes its learnable parameters.
The refined private latent codes $\hat{\textbf{Y}}_{\rm private}^{2}$ are then combined with the common latent codes $\textbf{Y}_{\rm common}^2$ to form the complete semantic representation $\textbf{S}_{2}$:
\begin{equation}
    \textbf{S}_2 = \left[\hat{\textbf{Y}}_{\rm private}^{2}, \textbf{Y}_{\rm common}^2\right].
\end{equation}
Finally, Eve inputs $\textbf{S}_{2}$ into the pre-trained Semantic StyleGAN generator to reconstruct the image:
\begin{equation}
    \hat{\textbf{X}}_{2} = f_{\mathrm{gen}}\left( \textbf{S}_{2} \right).
\end{equation}

\subsubsection{Overall Architecture}
The proposed secure SemCom system comprises three key components: 
(1) a pre-trained Semantic StyleGAN generator for extracting disentangled semantic representations and reconstructing source images, 
(2) NN-based DP protection and deprotection modules for privacy protection and recovery, and 
(3) a discriminator that assists in training the DP protection module by guiding it to generate noise distributions that closely approximate genuine DP noise, while ensuring that the noise can be mitigated at the legitimate receiver.
%
%
%
%
%
In the following, we detail each component and explain the calculation of the sensitivity $\Delta f$ in our proposed approximate DP protection mechanism.

\subsection{Pre-trained Semantic StyleGAN Generator}

We adopt the Semantic StyleGAN generator proposed in \cite{shi2022semanticstylegan} as both the encoder and decoder in our system, leveraging its bidirectional capability.
%
%
Specifically, in the \textit{forward direction}, the Semantic StyleGAN generator acts as a decoder, mapping the disentangled semantic representation to a reconstructed image at the receiver.
Conversely, in the \textit{reverse direction}, the generator functions as an encoder by performing GAN inversion at the transmitter, encoding the source image into a disentangled semantic representation.
We next introduce the details of the Semantic StyleGAN generator in both the forward and reverse directions.

%
%

\begin{figure}[t]
\begin{center}
\centerline{\includegraphics[width=1\linewidth]{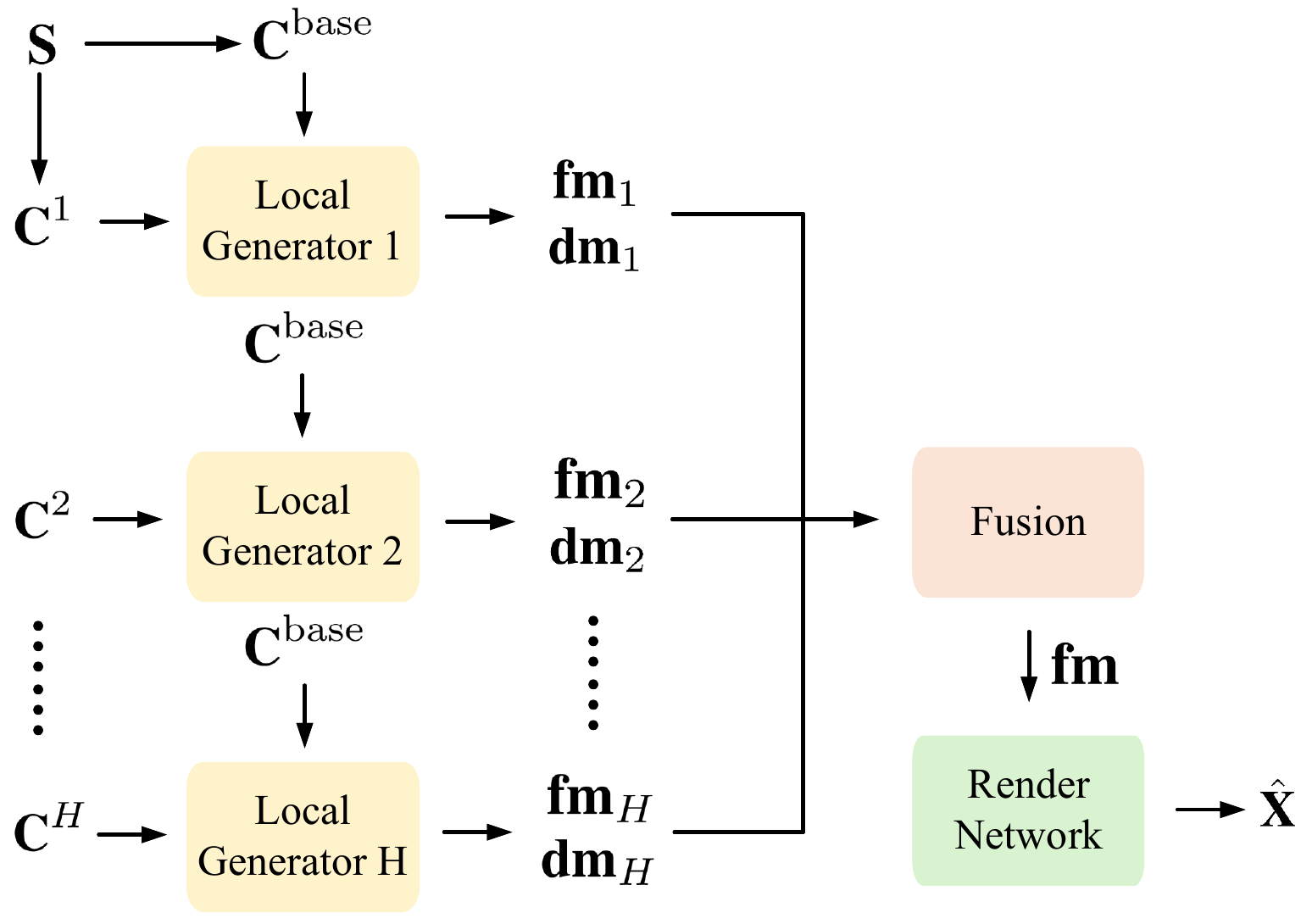}}
\caption{The framework of the Semantic StyleGAN generator.}
\label{styleGAN}
\end{center}
\vskip -0.3in
\end{figure}

The framework of the Semantic StyleGAN generator is illustrated in Fig.~\ref{styleGAN}.
At the core of the generator are $H$ local generators, each responsible for synthesizing a specific semantic area of the face image, such as the eyes, nose, or mouth. 
The input semantic representation, denoted by $\textbf{S}$, is divided into two parts: shared latent codes $\textbf{C}^{\rm base}$, which controls the coarse structure of the face, and $H$ local latent codes $\textbf{C}^{1}-\textbf{C}^{H}$, each controlling the shape and texture of its corresponding semantic area. 
Each local latent code is further decomposed into a shape code and a texture code.
The local generator corresponding to the $h$th semantic area takes as input both the shared latent codes and the $h$th local latent code. 
It is composed of modulated $1 \times 1$ convolutional layers and fully connected (FC) layers, which output a feature map $\textbf{fm}_{h}$ and a pseudo-depth map $\textbf{dm}_{h}$. 
The pseudo-depth maps are then used to generate a coarse segmentation mask $\textbf{m}$, which guides the fusion of all feature maps into an aggregated feature map $\textbf{fm}$. 
Finally, the aggregated feature map $\textbf{fm}$ is passed through a rendering network to produce the reconstructed image $\hat{\textbf{X}}$.



In contrast to the forward image generation process, the reverse direction, referred to as GAN inversion, aims to encode a given source image $\textbf{X}$ into a disentangled semantic representation $\textbf{Z}$ that effectively captures its semantic content in the latent space.
%
%
Formally, the objective of GAN inversion is to solve the following optimization problem:
\begin{equation}
    \arg\min_{\textbf{Z}} {\rm MSE} \left( \textbf{X}, f_{\rm gen}(\textbf{Z}) \right),
\end{equation}
where $f_{\rm gen}$ represents the generator, $f_{\rm gen}(\textbf{Z})$ represents the image generated from $\textbf{Z}$, and $\rm MSE$ represents the mean squared error (MSE).
The solution can be obtained through a fixed number of gradient descent iterations.

%

\subsection{NN-based DP Protection / Deprotection Modules}

\subsubsection{Discriminator and Adversarial Training}

%
%
Traditional DP protection methods are inherently non-invertible, which poses a significant challenge for reconstruction at the receiver. 
%
To overcome this limitation, we propose a novel NN-based DP protection module that generates DP noise with learnable pattern. 
This noise not only closely resembles genuine DP noise, thereby meeting privacy protection requirements, but also remains easily mitigated by the legitimate user through the corresponding NN-based DP deprotection module.

To protect the private latent codes using the NN-based DP protection module, 
the first step is to generate an intermediate variable, $\hat{\textbf{Z}}_{\rm private}^{1}$, which serves as a guiding representation during training.
Specifically, $\hat{\textbf{Z}}_{\rm private}^{1}$ is obtained by adding genuine DP noise to $\textbf{Z}_{\rm private}$: 
\begin{equation}
    \hat{\textbf{Z}}_{\rm private}^{1} = \textbf{Z}_{\rm private} + \textbf{n}_{\rm dp},
\end{equation}
where $\textbf{n}_{\rm dp} \sim Lap(0, \frac{\Delta f}{\epsilon})$, and the method for obtaining $\Delta f$ will be discussed in Subsection D. 
In the second step, the DP protection module perturbs the private latent codes under the guidance of $\hat{\textbf{Z}}_{\rm private}^{1}$, generating the perturbed latent codes $\hat{\textbf{Z}}_{\rm private}^{2}$.
%
%
%
It is important to emphasize that $\hat{\textbf{Z}}_{\rm private}^{1}$ serves only as an intermediate representation to guide the generation of $\hat{\textbf{Z}}_{\rm private}^{2}$ and is not transmitted.
%

%
%

\begin{figure}[t]
\begin{center}
\centerline{\includegraphics[width=1\linewidth]{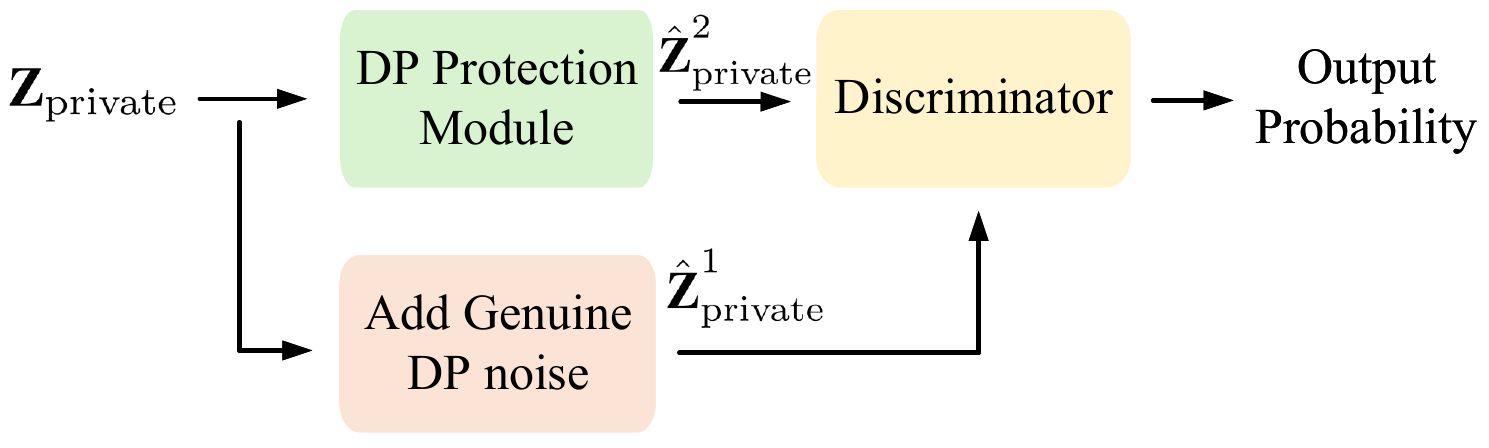}}
\caption{Illustration of the adversarial training process for the DP protection module.}
\label{adversarial framework}
\end{center}
\vskip -0.3in
\end{figure}

The DP protection module is trained in an adversarial manner with the help of a discriminator, as illustrated in Fig.~\ref{adversarial framework}.
The discriminator, which consists of two FC layers followed by a sigmoid output layer, is only used for the training of the DP protection module.
%
Specifically, it is trained to distinguish between  $\hat{\mathbf{Z}}^{1}_{\mathrm{private}}$ and $\hat{\mathbf{Z}}^{2}_{\mathrm{private}}$, 
while the DP protection module is optimized to generate outputs that the discriminator cannot reliably differentiate from genuinely DP-noised data.
%

%
The training objective for the discriminator $D(\cdot)$ is defined using the binary cross-entropy (BCE) loss:
\begin{equation}
    \mathcal{L}_{D} = -\mathbb{E}\left[\log D(\hat{\mathbf{Z}}^{1}_{\mathrm{private}})\right] - \mathbb{E}\left[\log(1 - D(\hat{\mathbf{Z}}^{2}_{\mathrm{private}}))\right].
\end{equation}
In contrast, the DP protection module $G(\cdot)$ is trained by minimizing the following loss:
\begin{equation}
    \mathcal{L}_{G} = \mathbb{E}\left[\log(1- D(\hat{\mathbf{Z}}^{2}_{\mathrm{private}}))\right],
\end{equation}
which encourages the DP protection module to generate perturbed latent codes whose noise patterns are indistinguishable from genuine DP noise. 
During training, the discriminator and the DP protection module are updated alternately. 
Moreover, the DP protection module must balance the trade-off between preserving reconstruction quality for the legitimate user and generating noise patterns that effectively resemble genuine DP noise. This trade-off will be further elaborated in Subsection E.

\begin{figure}[t]
\begin{center}
\centerline{\includegraphics[width=1\linewidth]{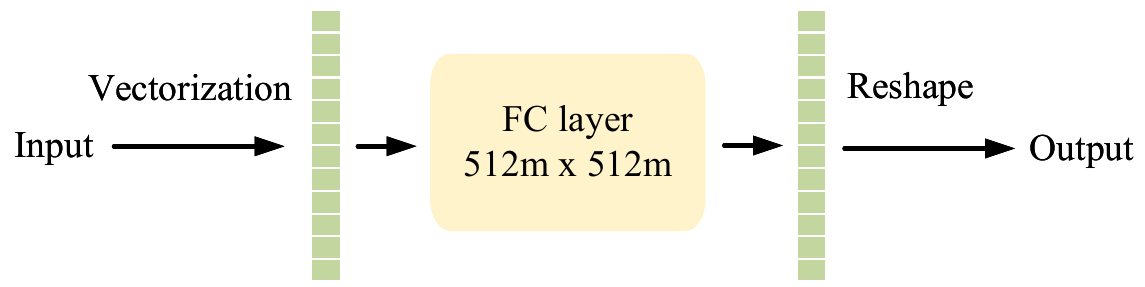}}
\caption{The network architecture of the proposed NN-based DP protection module.}
\label{DP pro}
\end{center}
\vskip -0.3in
\end{figure}

%
%

\subsubsection{Network Architectures}



The network architecture of the NN-based DP protection module is shown in Fig.~\ref{DP pro}.
%
%
%
This module first vectorizes the input private latent codes ${\textbf{Z}}_{\rm private} \in \mathbb{R}^{m \times 512}$ into a one-dimensional representation $\textbf{Z}_{\rm private}^{'} \in \mathbb{R}^{512m}$. 
The vectorized signal is then passed through the FC layer and reshaped to its original dimension to produce the output $\hat{\textbf{Z}}_{\rm private}^{2} \in \mathbb{R}^{m \times 512}$. 
The NN-based DP deprotection module adopts the same architecture as the DP protection module.
%
It vectorizes the input $\textbf{Y}_{\rm private}^1 / \textbf{Y}_{\rm private}^2 \in \mathbb{R}^{m \times 512}$, passes it through an FC layer, and then reshapes it back to the original dimension to obtain $\hat{\textbf{Y}}_{\rm private}^{1} / \hat{\textbf{Y}}_{\rm private}^{2} \in \mathbb{R}^{m \times 512}$.



\subsection{The Calculation of the Sensitivity}

%
In the context of DP for image protection \cite{xue2021dp}, 
the sensitivity $\Delta f$ quantifies the maximum difference between the latent codes of any two distinct images in the training dataset. 
Accordingly, in this paper, $\Delta f$ is formally defined as
\begin{equation}
   \Delta f \stackrel{.}{=} \sup_{\textbf{I}_1, \textbf{I}_2 \in \mathcal{D}} \left\| f_{\rm inv}(\textbf{I}_1) - f_{\rm inv}(\textbf{I}_2) \right\|_2,
   \label{sensi}
\end{equation}
where $\textbf{I}_1$ and $\textbf{I}_2$ are two different images sampled from the training dataset $\mathcal{D}$,
and $f_{\rm inv}$ represents the inversion process of the Semantic StyleGAN.
However, directly calculating $\Delta f$ using \eqref{sensi} presents two significant challenges. 
First, calculating the element-wise differences between the latent codes of all image pairs across a large training dataset results in considerable computational overhead. 
Second, certain source images may contain abnormal features that lead to outliers in the latent space, potentially distorting the sensitivity calculation.

To overcome these challenges, 
we adopt a clipping strategy that retains 99\% of the latent codes, effectively mitigating the effect of outliers and simultaneously reducing the computational cost of calculating $\Delta f$.
This process involves the following steps:
(1) First, transform each image in the training dataset into its corresponding latent codes.
(2) Next, determine the 0.5\% quantile $a$ and the 99.5\% quantile $b$ of all the latent code elements across the dataset.
(3) Finally, adjust any elements outside the range $\left[a, b\right]$ to either $a$ or $b$.

After clipping, the sensitivity $\Delta f$ is calculated as:
\begin{equation}
\begin{split}
& \Delta f = \left\| b\mathbb{I}_n - a\mathbb{I}_n \right\|_2 = \sqrt{(b - a)^{2} \cdot n},
\end{split}
\end{equation}
where $\mathbb{I}_n$ represents an all-ones vector of length $n$, and $n$ is the total number of elements in the latent codes for a single image.
This approach efficiently calculates $\Delta f$ while addressing the computational and statistical challenges associated with large datasets and outliers.

\subsection{Training Strategy}

In this subsection, we describe the training strategy employed for our system under two different settings for the eavesdropper.

\subsubsection{Proposed System (Basic Eavesdropper)}

In the basic eavesdropper setting, the eavesdropper is unaware that DP noise with learnable pattern has been added to the private latent codes. 
Only the legitimate network (DP protection / deprotection modules of the legitimate user) and the discriminator are trained. 
The loss function for training the discriminator can be expressed as:
\begin{equation}
    \mathcal{L}_{(1)} = - \mathbb{E}\left[\log D(\hat{\mathbf{Z}}^{1}_{\mathrm{private}})\right] - \mathbb{E}\left[\log(1 - D(\hat{\mathbf{Z}}^{2}_{\mathrm{private}}))\right].
\end{equation}
The loss function for training the legitimate network can be expressed as:
\begin{equation}
    \mathcal{L}_{(2)} = {\rm MSE} \left( {\textbf{Z},\textbf{S}_{1}} \right) + \lambda \cdot \mathbb{E}\left[\log(1- D(\hat{\mathbf{Z}}^{2}_{\mathrm{private}}))\right].
    \label{eq33}
\end{equation}
The first term of $\mathcal{L}_{(2)}$ measures the image reconstruction performance of the legitimate user. 
The second term of $\mathcal{L}_{(2)}$ measures how closely the DP noise with learnable pattern resembles the genuine DP noise, with $\lambda$ serving as the trade-off hyperparameter to balance these objectives.
The two loss functions, $\mathcal{L}_{(1)}$ and $\mathcal{L}_{(2)}$, are optimized alternately. 

\subsubsection{Proposed System (Stronger Eavesdropper)}

In the stronger eavesdropper setting, the eavesdropper is aware that the semantic representation has been perturbed using DP noise with learnable pattern. 
The eavesdropper also has access to the network architecture of the DP deprotection module but does not know which latent codes have been protected. 
%
%
Unlike existing secure SemCom methods, in this setting, the performance of the eavesdropper is not considered during the training of the legitimate network (DP protection / deprotection modules), since the eavesdropper typically does not cooperate with the legitimate user. 
To simulate a realistic adversarial scenario, we allow the eavesdropper to train her network after the legitimate network has been trained.
%
Specifically, we adopt a two-stage training strategy.

\textit{First Training Stage:} 
This training stage follows the same procedure as that of the basic eavesdropper setting.

%

\textit{Second Training Stage:} In this stage, only the eavesdropper's network (DP deprotection module of the eavesdropper) is trained. The loss function for this stage is:
\begin{equation}
    \mathcal{L} = {\rm MSE} \left( {\textbf{Z},\textbf{S}_{2}} \right),
\end{equation}
which measures the image reconstruction performance of the eavesdropper.

\section{Performance Evaluation}

\subsection{Experimental Settings}

\subsubsection{Pre-trained Model and Dataset}
We employ the pre-trained Semantic StyleGAN model trained on the CelebAMask-HQ dataset \cite{lee2020maskgan}, where each image resized to $512 \times 512$. 
For the proposed system, we utilize the CelebAMask-HQ dataset for both training and testing. 
This dataset consists of 30,000 $1024 \times 1024$ RGB face images, with the first 28,000 images used for training and the remaining 2,000 images used for testing.
All images are resized to $512 \times 512$ to maintain consistency with the pre-trained model.

\subsubsection{Private Latent Codes}

The latent codes of each image have a dimension of $28 \times 512$, where 28 represents the number of latent codes.
Among these, the first two latent codes are shared latent codes, while the remaining latent codes are categorized as shape and texture codes (local latent codes). 
Each latent code has a length of 512. 
Thus, $H = (28-2)/2 = 13$, where $H$ represents the number of groups of local latent codes.
Shared latent codes are always private latent codes. 
Additionally, certain latent codes from the local latent codes are selected as private latent codes.

\begin{figure}[t]
\begin{center}
\centerline{\includegraphics[width=0.95\linewidth]{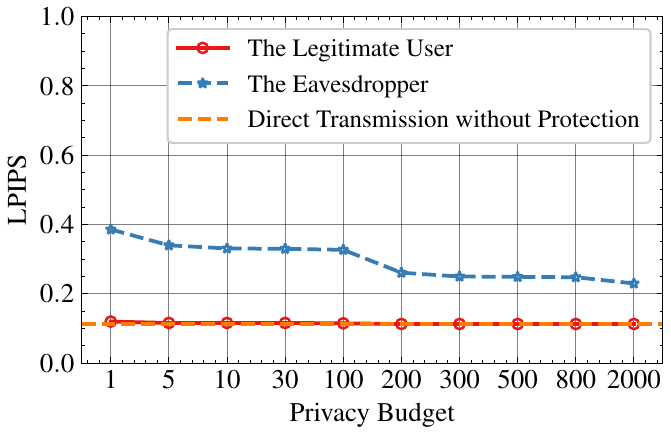}}
\caption{The LPIPS performance of the \textit{Proposed System (Basic Eavesdropper)} under different privacy budgets $\epsilon$, where the channel SNR is set to 20 dB.}
\label{fig.LPIPS1}
\end{center}
\vskip -0.3in
\end{figure}

\begin{figure}[t]
\begin{center}
\centerline{\includegraphics[width=0.95\linewidth]{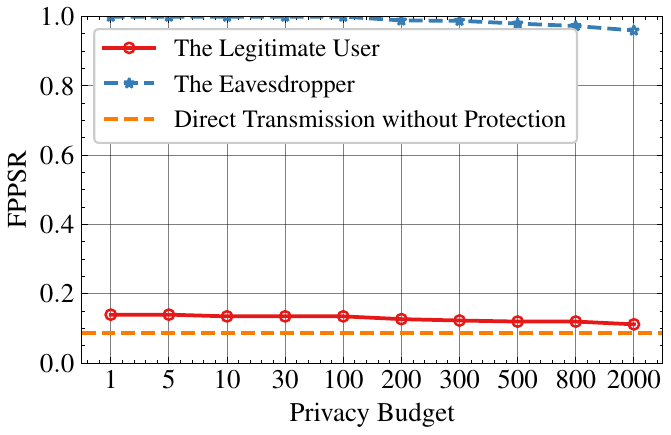}}
\caption{The FPPSR performance of the \textit{Proposed System (Basic Eavesdropper)} under different privacy budgets $\epsilon$, where the channel SNR is set to 20 dB.}
\label{fig.FPPSR1}
\end{center}
\vskip -0.3in
\end{figure}

\subsubsection{Privacy Budget and Channel SNR}
The sensitivity value $\Delta f$ is 351.88, and the privacy budget $\epsilon$ takes values from the set $\{1,5,10,30,100,200,300,500,800,2000\}$. 
In this paper, we assume that the eavesdropper experiences the same channel conditions as the legitimate user, which represents the most challenging scenario for the legitimate user.
Therefore, $\sigma_{1} = \sigma_{2}$, and $\rm SNR= \rm SNR_{\rm leg} = \rm SNR_{\rm eve}$.
The channel SNR values are selected from the set $\{0,5,10,15,20\}$ dB. 

\subsubsection{Training Settings}
Our experiments are conducted on a single NVIDIA RTX A6000 GPU.
The batch size is set to 256, and the cosine annealing warm restarts optimizer is used to train the proposed system. 

For the \textit{Proposed System (Basic Eavesdropper)}, $\lambda$ is set to $1 \times 10^{-3}$. 
The shared latent codes, as well as the 4th to 7th latent codes, are selected as private latent codes. 
The initial learning rate is $3 \times 10^{-4}$, and the training lasts for 100 epochs.

For the \textit{Proposed System (Stronger Eavesdropper)}, $\lambda$ is set to $1 \times 10^{-3}$.  
The shared latent codes and the 4th to 13th latent codes are private latent codes for the legitimate user. The eavesdropper guesses the shared latent codes and the 6th to 7th latent codes are private latent codes.
In the first training stage, the initial learning rate is $3 \times 10^{-4}$, and the training lasts for 100 epochs. 
In the second training stage, the initial learning rate is $3 \times 10^{-4}$, and the training lasts for 50 epochs.

\subsubsection{The Benchmarks}


To evaluate the effectiveness of our proposed framework, we compare it against two representative benchmark methods.


%
In the first benchmark, we focus on scenarios without any security mechanisms.
The latent codes $\textbf{Z}$ are directly transmitted from Alice to Bob without employing the DP protection module. 
Both Bob and Eve reconstruct the source image directly based on the received latent codes $\textbf{Y}_1$ and $\textbf{Y}_2$. 
Since they share the same channel SNR, their image reconstruction performances are identical. 
We refer to this benchmark as \textit{Direct Transmission without Protection}. 
It is used to evaluate the effectiveness of privacy protection in our proposed system and its effect on the legitimate user's task performance.

The second benchmark, referred to as \textit{Traditional DP Protection} \cite{li2021differentially}, was proposed by Li \textit{et al.}
In this benchmark, we directly apply standard DP mechanisms \cite{dwork2006differential,li2021differentially} to the private latent codes by adding Laplace noise calibrated according to a specified privacy budget. 
At the receiver, we train a dedicated NN that learns to remove the DP noise. 
This benchmark provides a formal privacy guarantee and serves as a baseline to evaluate the effectiveness of our proposed method in two aspects: its ability to mitigate the degradation in task performance caused by the non-invertibility of DP noise, and its capacity to provide effective image protection under DP constraints.

\begin{figure}[t]
\begin{center}
\centerline{\includegraphics[width=0.95\linewidth]{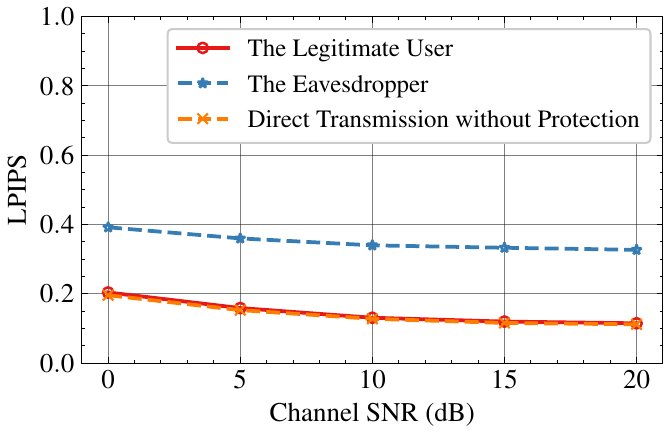}}
\caption{The LPIPS performance of the \textit{Proposed System (Basic Eavesdropper)} under different channel SNRs, where the privacy budget $\epsilon$ is set to 100.}
\label{fig.SNR_LPIPS}
\end{center}
\vskip -0.3in
\end{figure}

\begin{figure}[t]
\begin{center}
\centerline{\includegraphics[width=0.95\linewidth]{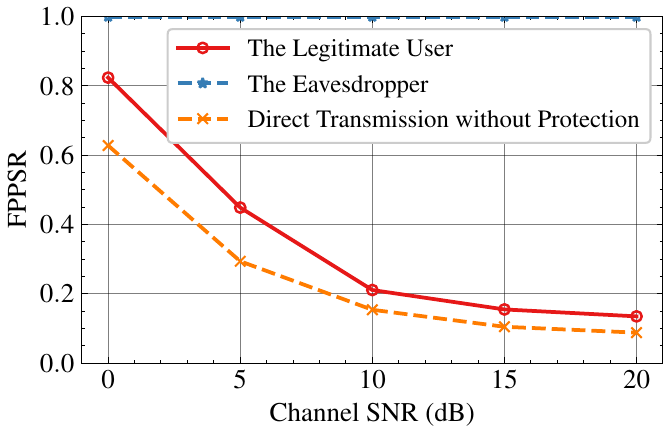}}
\caption{The FPPSR performance of the \textit{Proposed System (Basic Eavesdropper)} under different channel SNRs, where the privacy budget $\epsilon$ is set to 100.}
\label{fig.SNR_FPPSR}
\end{center}
\vskip -0.3in
\end{figure}

\begin{figure*}[t]
\centering
\subfigure[Source image]{
\begin{minipage}[t]{0.16\linewidth}
\centering
\includegraphics[width=0.9in]{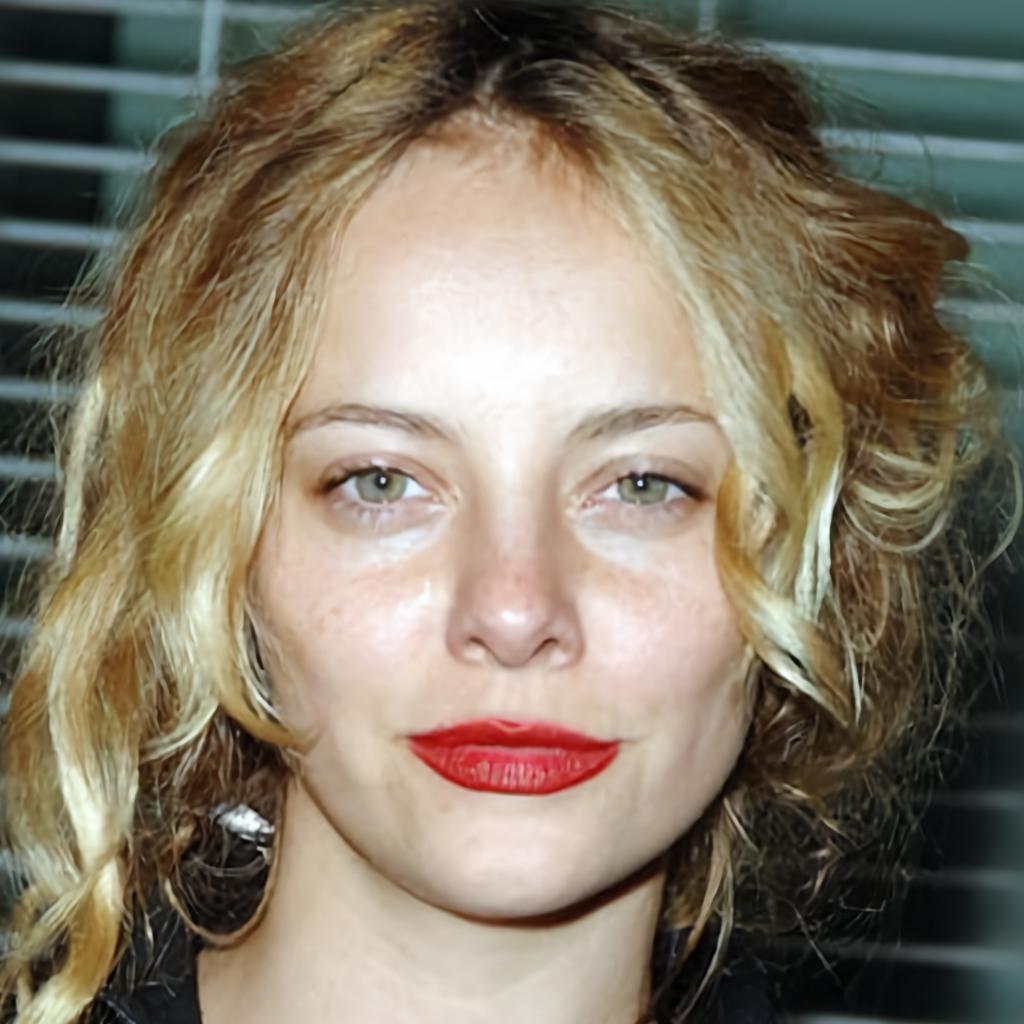}
\end{minipage}%
}%
\subfigure[$\epsilon = 1$]{
\begin{minipage}[t]{0.16\linewidth}
\centering
\includegraphics[width=0.9in]{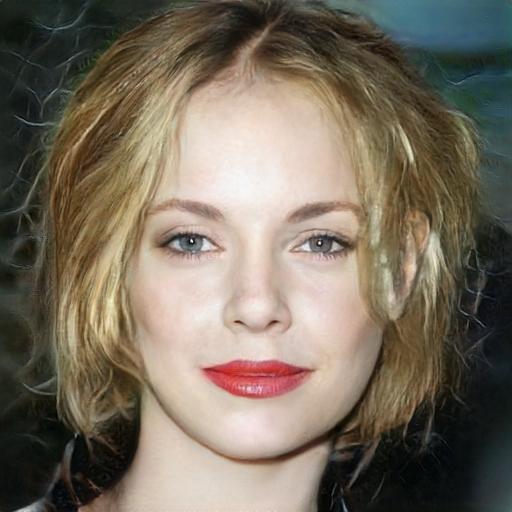}
\end{minipage}%
}%
\subfigure[$\epsilon = 5$]{
\begin{minipage}[t]{0.16\linewidth}
\centering
\includegraphics[width=0.9in]{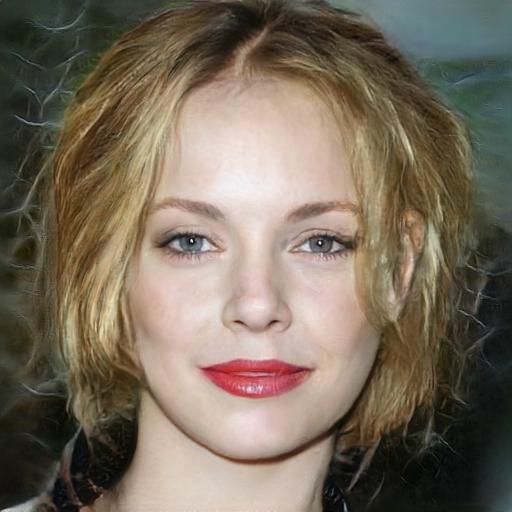}
\end{minipage}
}%
\subfigure[$\epsilon = 10$]{
\begin{minipage}[t]{0.16\linewidth}
\centering
\includegraphics[width=0.9in]{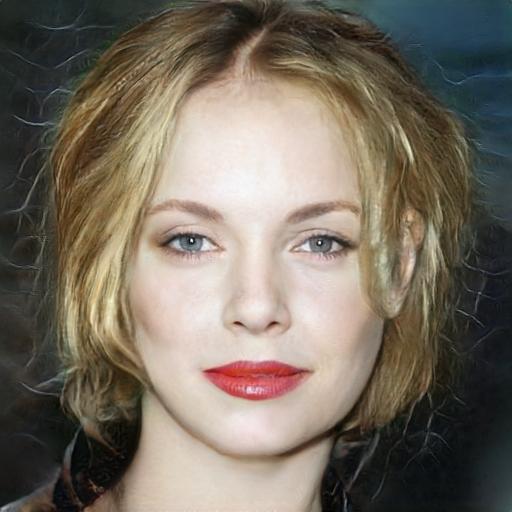}
\end{minipage}
}%
\subfigure[$\epsilon = 30$]{
\begin{minipage}[t]{0.16\linewidth}
\centering
\includegraphics[width=0.9in]{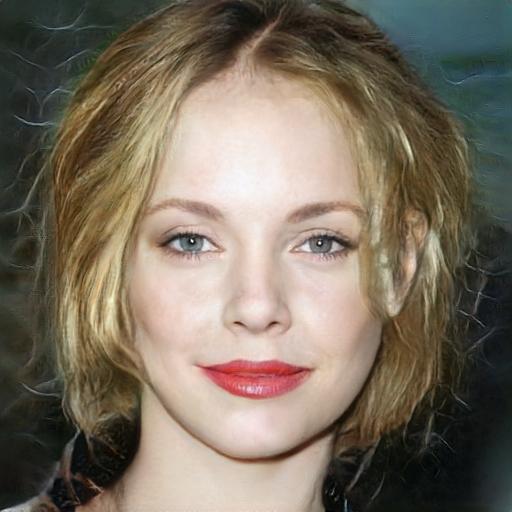}
\end{minipage}
}%
\subfigure[$\epsilon = 100$]{
\begin{minipage}[t]{0.16\linewidth}
\centering
\includegraphics[width=0.9in]{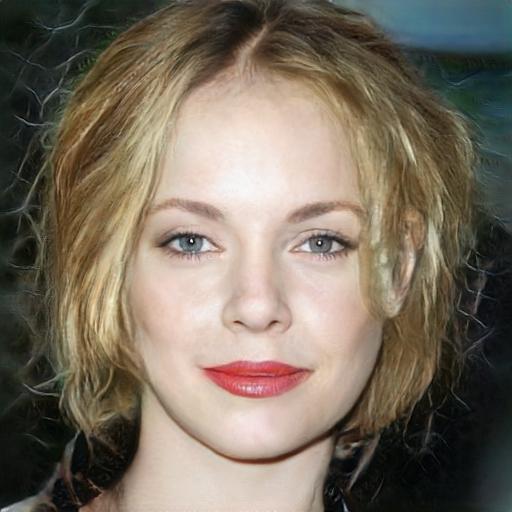}
\end{minipage}
}%
\\
\subfigure[$\epsilon = 200$]{
\begin{minipage}[t]{0.16\linewidth}
\centering
\includegraphics[width=0.9in]{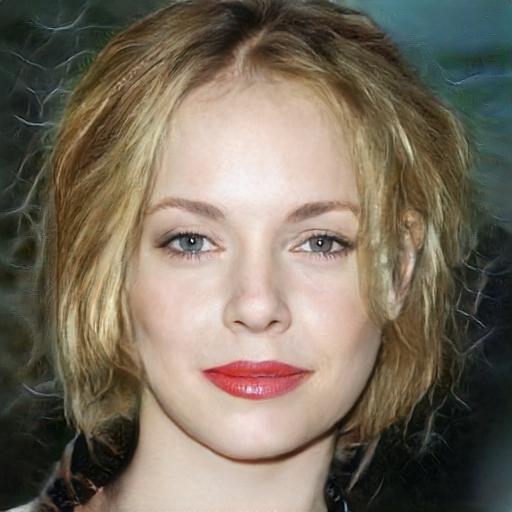}
\end{minipage}
}%
\subfigure[$\epsilon = 300$]{
\begin{minipage}[t]{0.16\linewidth}
\centering
\includegraphics[width=0.9in]{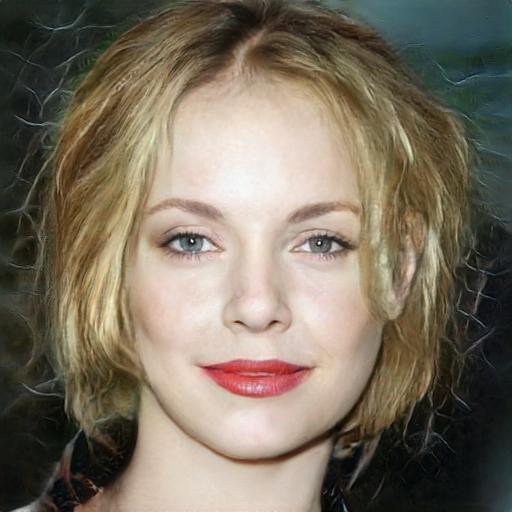}
\end{minipage}
}%
\subfigure[$\epsilon = 500$]{
\begin{minipage}[t]{0.16\linewidth}
\centering
\includegraphics[width=0.9in]{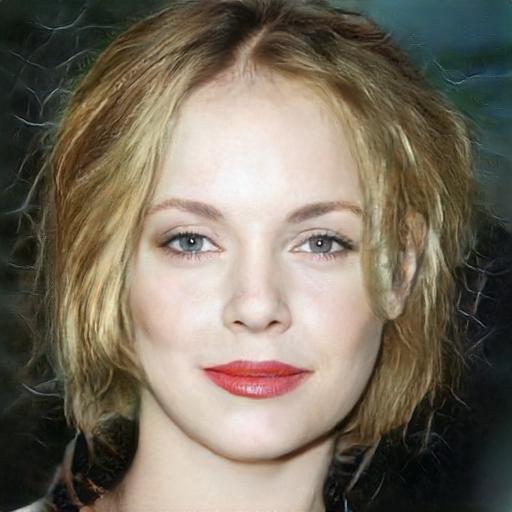}
\end{minipage}
}%
\subfigure[$\epsilon = 800$]{
\begin{minipage}[t]{0.16\linewidth}
\centering
\includegraphics[width=0.9in]{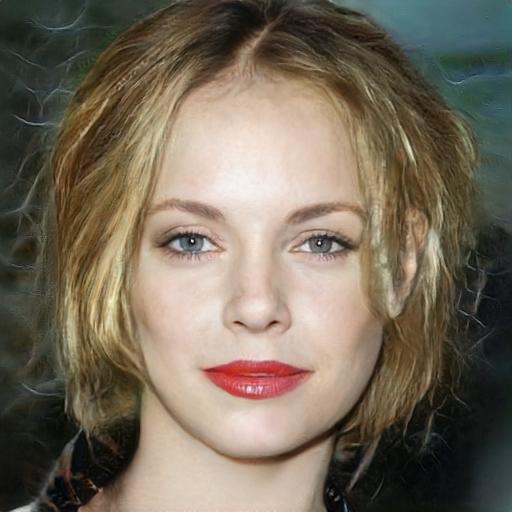}
\end{minipage}
}%
\subfigure[$\epsilon = 2000$]{
\begin{minipage}[t]{0.16\linewidth}
\centering
\includegraphics[width=0.9in]{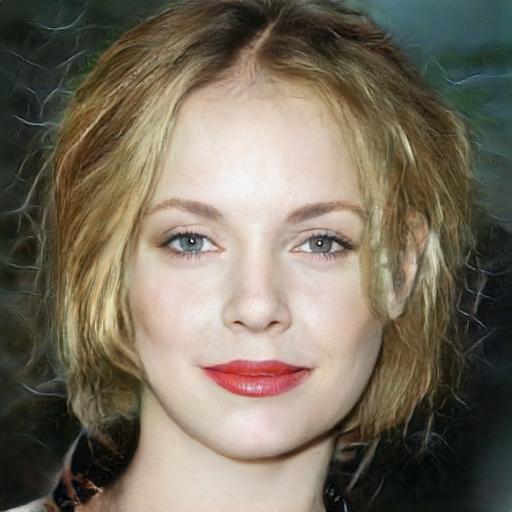}
\end{minipage}
}%
\centering
\caption{Visual analysis of the reconstructed images by the legitimate user under different privacy budgets $\epsilon$ in the \textit{Proposed System (Basic Eavesdropper)}. The channel SNR is set to 20 dB.}
\label{fig.visual1}
\vskip -0.1in
\end{figure*}

\begin{figure*}[t]
\centering
\subfigure[Source image]{
\begin{minipage}[t]{0.16\linewidth}
\centering
\includegraphics[width=0.9in]{with_dp_only_2/original_image.jpg}
\end{minipage}%
}%
\subfigure[$\epsilon = 1$]{
\begin{minipage}[t]{0.16\linewidth}
\centering
\includegraphics[width=0.9in]{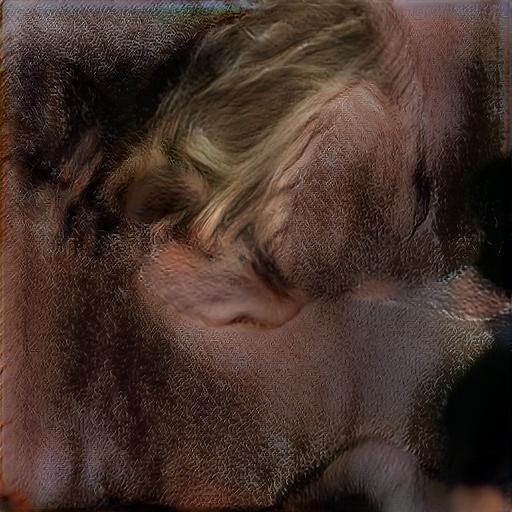}
\end{minipage}%
}%
\subfigure[$\epsilon = 5$]{
\begin{minipage}[t]{0.16\linewidth}
\centering
\includegraphics[width=0.9in]{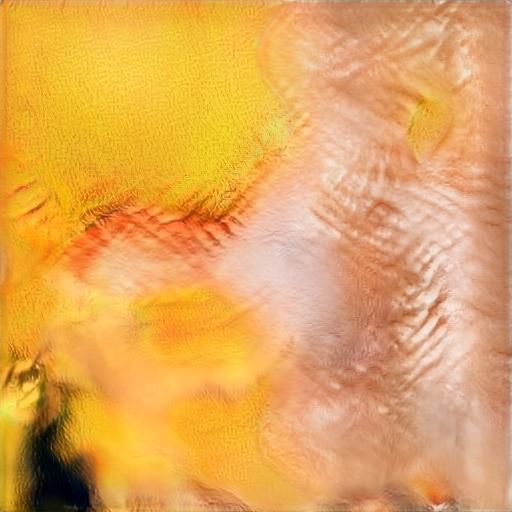}
\end{minipage}
}%
\subfigure[$\epsilon = 10$]{
\begin{minipage}[t]{0.16\linewidth}
\centering
\includegraphics[width=0.9in]{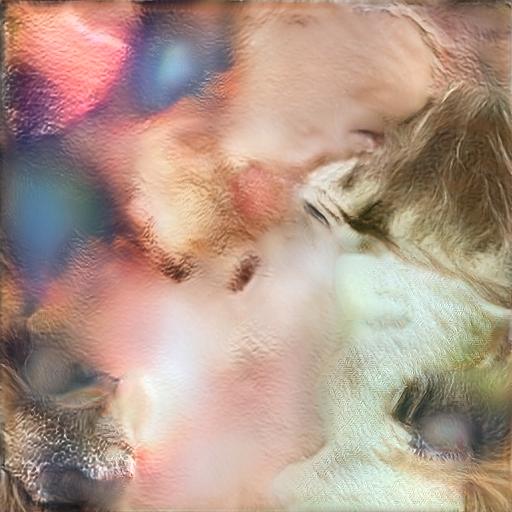}
\end{minipage}
}%
\subfigure[$\epsilon = 30$]{
\begin{minipage}[t]{0.16\linewidth}
\centering
\includegraphics[width=0.9in]{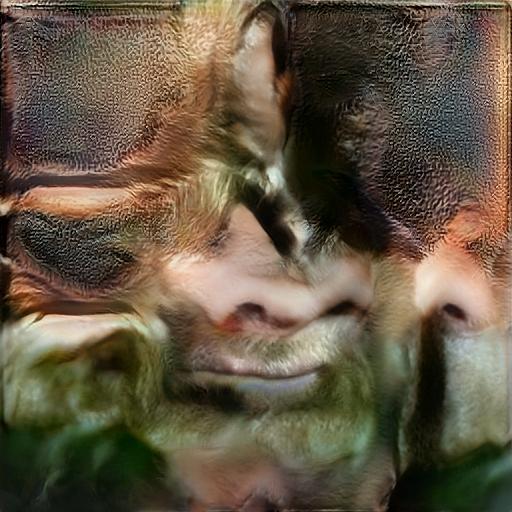}
\end{minipage}
}%
\subfigure[$\epsilon = 100$]{
\begin{minipage}[t]{0.16\linewidth}
\centering
\includegraphics[width=0.9in]{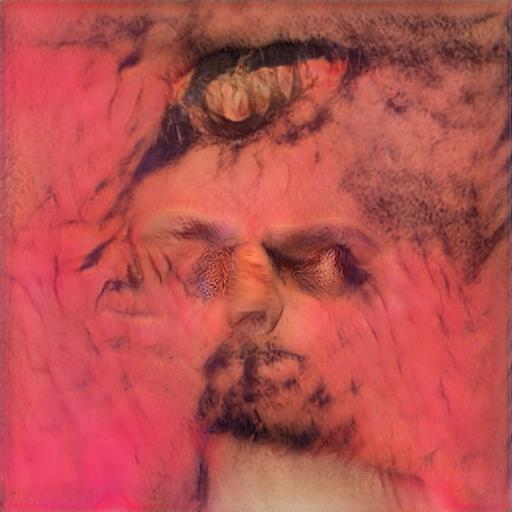}
\end{minipage}
}%
\\
\subfigure[$\epsilon = 200$]{
\begin{minipage}[t]{0.16\linewidth}
\centering
\includegraphics[width=0.9in]{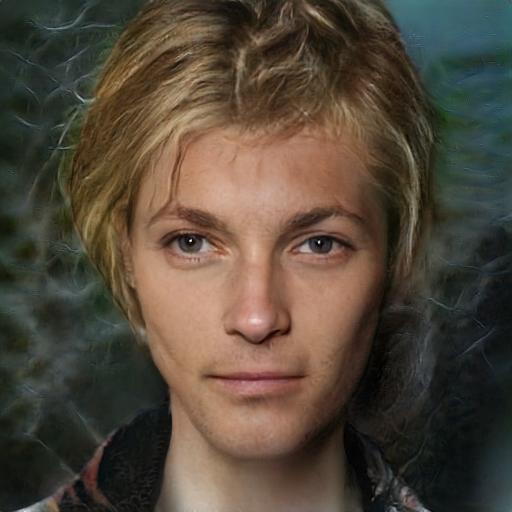}
\end{minipage}
}%
\subfigure[$\epsilon = 300$]{
\begin{minipage}[t]{0.16\linewidth}
\centering
\includegraphics[width=0.9in]{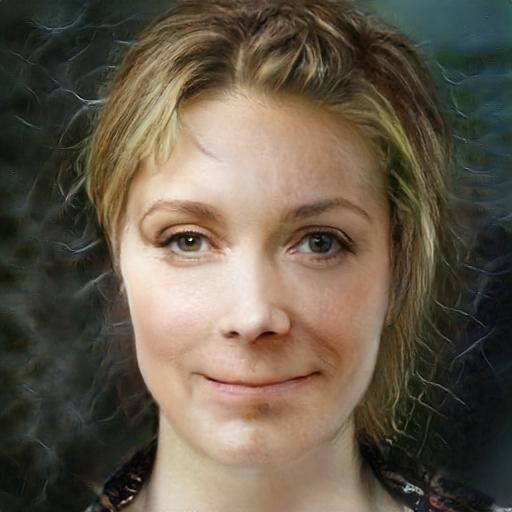}
\end{minipage}
}%
\subfigure[$\epsilon = 500$]{
\begin{minipage}[t]{0.16\linewidth}
\centering
\includegraphics[width=0.9in]{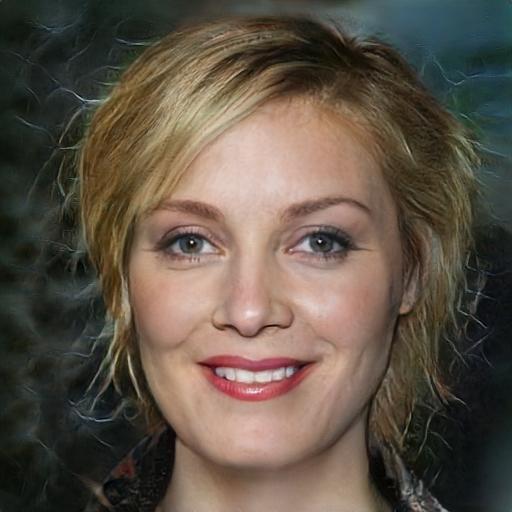}
\end{minipage}
}%
\subfigure[$\epsilon = 800$]{
\begin{minipage}[t]{0.16\linewidth}
\centering
\includegraphics[width=0.9in]{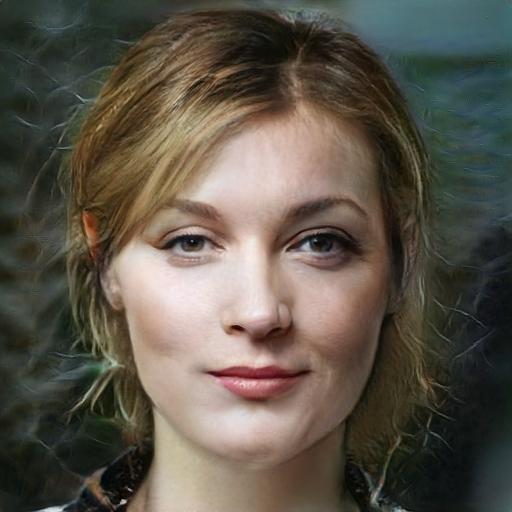}
\end{minipage}
}%
\subfigure[$\epsilon = 2000$]{
\begin{minipage}[t]{0.16\linewidth}
\centering
\includegraphics[width=0.9in]{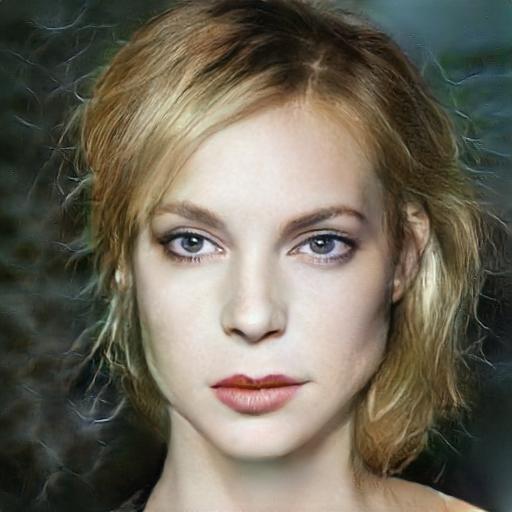}
\end{minipage}
}%
\centering
\caption{Visual analysis of the reconstructed images by the eavesdropper under different privacy budgets $\epsilon$ in the \textit{Proposed System (Basic Eavesdropper)}. The channel SNR is set to 20 dB.}
\label{fig.visual2}
\vskip -0.1in
\end{figure*}


\subsection{Proposed System under the Basic Eavesdropper Setting}


\subsubsection{Different Privacy Budgets}

Fig.~\ref{fig.LPIPS1} shows the LPIPS performance under different privacy budgets $\epsilon$, with the channel SNR fixed at 20 dB. 
The LPIPS metric quantifies perceptual similarity, where a lower value indicates better image reconstruction performance, and a higher value for the eavesdropper represents improved system security.
Across the entire range of privacy budgets, our proposed method ensures that the legitimate user's LPIPS remains low and close to the benchmark value of 0.112, reflecting minimal degradation in reconstruction quality for the legitimate user.
When $\epsilon = 1$, the LPIPS for the legitimate user is 0.120. As $\epsilon$ increases, the value decreases and stabilizes at 0.113 when $\epsilon \geq 200$, which is almost identical to \textit{Direct Transmission without Protection}.
In contrast, the LPIPS for the eavesdropper is 0.386 at $\epsilon = 1$, indicating severe reconstruction distortion. Although it gradually decreases as $\epsilon$ increases, it remains around 0.230 at $\epsilon = 2000$, which is considerably higher than that of the legitimate user.
This highlights the ability of the DP protection module to degrade the reconstruction quality for the eavesdropper, thereby enhancing system security.

Fig.~\ref{fig.FPPSR1} shows the FPPSR performance under the same conditions. 
FPPSR reflects the system's ability to distinguish reconstructed face images from the original ones. 
Lower FPPSR values for the legitimate user indicate better image reconstruction quality, while higher FPPSR values for the eavesdropper imply stronger privacy protection.
%
For the legitimate user, FPPSR begins at 0.140 when $\epsilon = 1$ and gradually declines to 0.112 at $\epsilon = 2000$. This value closely approaches the benchmark of 0.088.
%
The eavesdropper consistently exhibits high FPPSR values across all privacy budgets. Starting at 1.0 when $\epsilon = 1$, the value only slightly decreases to 0.960 at $\epsilon = 2000$, indicating that the eavesdropper's reconstructed images are still highly distorted, unrecognizable, or slightly misleading.
These results highlight the robustness of the \textit{Proposed System (Basic Eavesdropper)} in achieving a strong balance between preserving reconstruction quality for the legitimate user and ensuring privacy protection against adversaries.


\begin{figure}[t]
\begin{center}
\centerline{\includegraphics[width=0.95\linewidth]{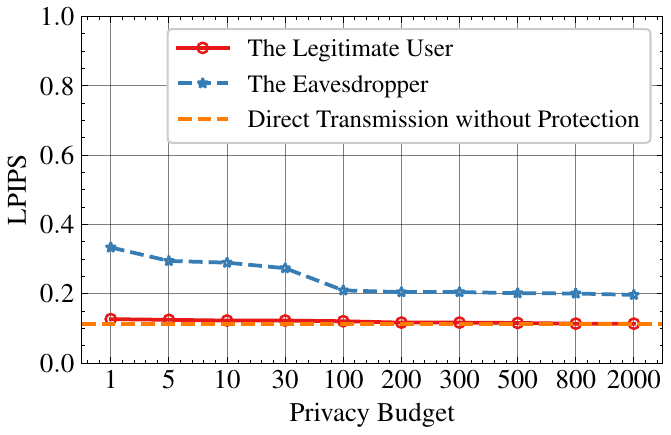}}
\caption{The LPIPS performance of the \textit{Proposed System (Stronger Eavesdropper)} under different privacy budgets $\epsilon$, where the channel SNR is set to 20 dB.}
\label{fig.LPIPS2}
\end{center}
\vskip -0.3in
\end{figure}

\begin{figure}[t]
\begin{center}
\centerline{\includegraphics[width=0.95\linewidth]{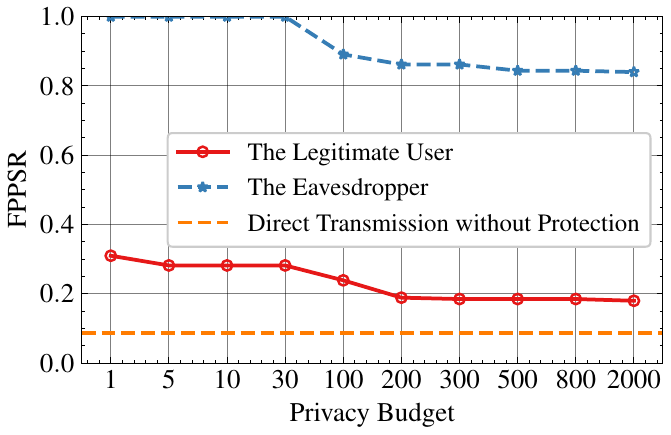}}
\caption{The FPPSR performance of the \textit{Proposed System (Stronger Eavesdropper)} under different privacy budgets $\epsilon$, where the channel SNR is set to 20 dB.}
\label{fig.FPPSR2}
\end{center}
\vskip -0.3in
\end{figure}

\begin{figure}[t]
\begin{center}
\centerline{\includegraphics[width=0.95\linewidth]{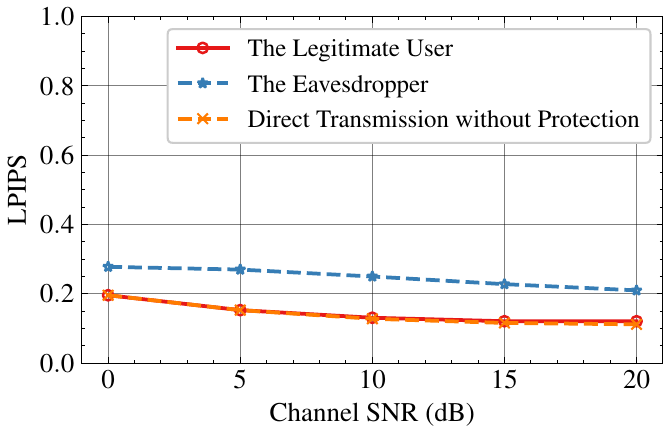}}
\caption{The LPIPS performance of the \textit{Proposed System (Stronger Eavesdropper)} under different channel SNRs, where the privacy budget $\epsilon$ is set to 100.}
\label{fig.SNR_LPIPS2}
\end{center}
\vskip -0.3in
\end{figure}

\begin{figure}[t]
\begin{center}
\centerline{\includegraphics[width=0.95\linewidth]{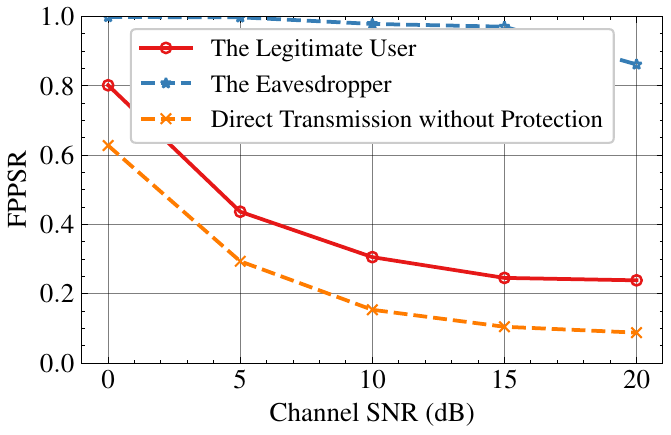}}
\caption{The FPPSR performance of the \textit{Proposed System (Stronger Eavesdropper)} under different channel SNRs, where the privacy budget $\epsilon$ is set to 100.}
\label{fig.SNR_FPPSR2}
\end{center}
\vskip -0.3in
\end{figure}


\subsubsection{Different Channel SNRs}

Fig.~\ref{fig.SNR_LPIPS} shows the LPIPS performance over different channel SNRs, while the privacy budget $\epsilon$ is fixed at 100. We can see that our method achieves comparable reconstruction quality to \textit{Direct Transmission without Protection} for the legitimate user across all channel SNRs, with LPIPS values only slightly higher than the benchmark. 
In contrast, the LPIPS values for the eavesdropper remain significantly higher than the benchmark at all channel SNRs, reflecting degraded perceptual similarity and thus enhanced system security. 
Specifically, at a channel SNR of 20 dB, the eavesdropper's LPIPS is 0.327, which is 0.212 higher than that of the legitimate user and 0.215 higher than the benchmark.

Fig.~\ref{fig.SNR_FPPSR} shows the FPPSR performance under the same settings. Again, the legitimate user's FPPSR performance closely follows the benchmark and improves as SNR increases.
%
Meanwhile, the eavesdropper's FPPSR remains consistently at 1.0 across all SNRs. 
This suggests that the eavesdropper consistently fails to reconstruct images that resemble the original identity, regardless of the channel quality.
In conclusion, our method effectively maintains high task performance while substantially impairing the eavesdropper's ability to recover recognizable images, across all channel SNRs. 

\subsubsection{Visual Evaluation}

Fig.~\ref{fig.visual1} and Fig.~\ref{fig.visual2} provide visual comparisons of the reconstructed images by the legitimate user and the eavesdropper under different privacy budgets $\epsilon$, with the channel SNR fixed at 20 dB. 
In Fig.~\ref{fig.visual1}, we observe the reconstruction results for the legitimate user across a range of privacy budgets. 
%
%
When $\epsilon = 1$, the reconstructed image already exhibits a high degree of semantic consistency with the source.
As $\epsilon$ increases, the visual similarity further improves.
When $\epsilon \geq 100$, 
the reconstructed images closely resemble the source image, demonstrating the robust deprotection capability of our DP deprotection module. 
These results indicate minimal perceptual differences and high-quality reconstruction.

In Fig.~\ref{fig.visual2}, we show the reconstructed results of the eavesdropper under the same range of privacy budgets. When $\epsilon$ is small (e.g., $\epsilon = 1, 5, 10, 30, 100$), the reconstructed images appear highly chaotic, lacking coherent facial structure, and are virtually unrecognizable.
This indicates that our method effectively obfuscates sensitive semantic representation at low privacy budgets. As $\epsilon$ increases (e.g., $\epsilon = 200, 300, 500, 800, 2000$), the eavesdropper is able to recover structured face images.
However, even at these higher privacy levels, the reconstructed images still exhibit significant deviations from the source image, making it visually implausible to infer the true identity.
These results highlight the flexibility of the proposed approach, which can achieve varying levels of privacy protection by adjusting the privacy budget. 
By carefully selecting the privacy budget, our proposed system can either effectively protect private information or generate fake yet artifact-mitigated images to slightly mislead the eavesdropper, ensuring robust system security while maintaining high reconstruction quality for the legitimate user.
%


\subsection{Proposed System under the Stronger Eavesdropper Setting}


\subsubsection{Different Privacy Budgets}

Fig.~\ref{fig.LPIPS2} presents the LPIPS performance of the \textit{Proposed System (Stronger Eavesdropper)} under different privacy budgets, with the channel SNR fixed at 20 dB. 
As shown in Fig.~\ref{fig.LPIPS2}, the LPIPS values for the legitimate user remain consistently low across all privacy budgets, indicating strong reconstruction fidelity. Specifically, the LPIPS value begins at 0.127 when $\epsilon = 1$ and decreases to 0.114 as $\epsilon$ increases to 2000. These values are consistently close to the benchmark of 0.112, demonstrating that the NN-based DP deprotection module effectively mitigates the effect of the added DP noise with learnable pattern.
In contrast, the LPIPS values for the eavesdropper remain higher than those for the legitimate user. 
When $\epsilon = 1$, the eavesdropper's LPIPS is 0.334, suggesting severely distorted reconstructions. 
Even as $\epsilon$ increases, the LPIPS decreases slowly and reaches 0.197 at $\epsilon = 2000$, which is still considerably higher than the legitimate user's LPIPS. 
This performance gap verifies the robustness of the proposed method in preserving semantic fidelity for the legitimate user while effectively preventing the eavesdropper from accessing private information.

Fig.~\ref{fig.FPPSR2} shows the corresponding FPPSR performance under varying $\epsilon$. For the legitimate user, the FPPSR decreases from 0.310 at $\epsilon = 1$ to 0.180 at $\epsilon = 2000$, gradually approaching the benchmark value of 0.088. This trend is consistent with the LPIPS results, indicating that the legitimate user can reconstruct high-quality images with increasing privacy budgets.
On the other hand, the eavesdropper consistently exhibits very high FPPSR values, starting at 1.0 when $\epsilon = 1$ and decreasing to 0.840 at $\epsilon = 2000$. 
These results suggest that the eavesdropper's reconstructed images are either chaotic or substantially different from the original identity.
The substantial FPPSR gap across all $\epsilon$ values further demonstrates the effectiveness of the proposed system.



\subsubsection{Different Channel SNRs}


Fig.~\ref{fig.SNR_LPIPS2} illustrates the LPIPS performance of the \textit{Proposed System (Stronger Eavesdropper)} under different channel SNRs, with the privacy budget $\epsilon$ fixed at 100. 
As shown in Fig.~\ref{fig.SNR_LPIPS2}, the LPIPS values for the legitimate user decrease from 0.196 at 0 dB to 0.121 at 20 dB, indicating a consistent improvement in perceptual reconstruction quality as the channel conditions improve. 
Notably, the LPIPS values closely approach the benchmark curve, demonstrating that the NN-based DP deprotection module effectively mitigates the effect of the added DP noise with learnable pattern and preserves high semantic fidelity for the legitimate user, even in low SNR regimes.
In contrast, the eavesdropper consistently exhibits higher LPIPS values than the legitimate user across all SNR levels, starting at 0.278 when SNR is 0 dB and decreasing to 0.210 at 20 dB. 
Although the LPIPS values for the eavesdropper slightly decrease with increasing channel SNR, they consistently exceed those of the legitimate user by a large margin, confirming that the proposed method significantly degrades the perceptual similarity of the eavesdropper's reconstructed images. 
This performance gap highlights the proposed system's robustness in resisting unauthorized reconstruction, 
owing to the DP protection module and the knowledge gap, even when the eavesdropper has channel conditions equivalent to the legitimate user.

Fig.~\ref{fig.SNR_FPPSR2} presents the FPPSR performance under the same conditions as in Fig.~\ref{fig.SNR_LPIPS2}. 
As shown in Fig.~\ref{fig.SNR_FPPSR2}, the FPPSR for the legitimate user decreases significantly from 0.802 at 0 dB to 0.239 at 20 dB.  
This trend is consistent with the LPIPS results and further confirms that the preservation of facial identity improves as channel quality increases.
%
Notably, the FPPSR values for the legitimate user remain close to the benchmark, reflecting strong identity retention capabilities.
In contrast, the eavesdropper exhibits consistently high FPPSR values across all channel SNRs, beginning at 0.999 and reducing only to 0.862 at 20 dB. 
These elevated values indicate that the eavesdropper's reconstructed images are largely misidentified or appear chaotic, effectively obscuring the original facial identity. 
The persistent FPPSR gap further validates the effectiveness of the proposed system in protecting private information.

\begin{figure}[t]
\begin{center}
\centerline{\includegraphics[width=0.95\linewidth]{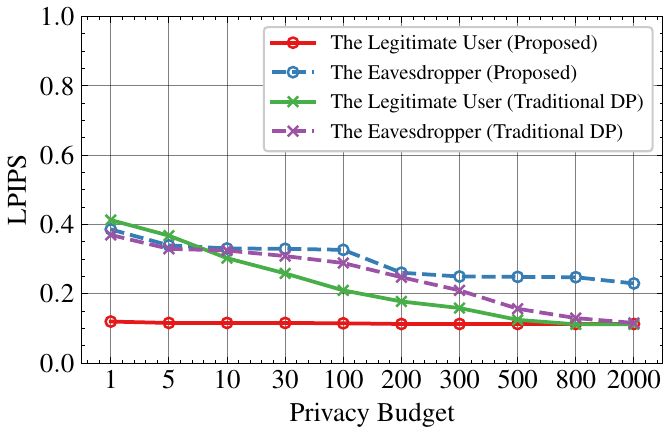}}
\caption{The LPIPS performance of the \textit{Proposed System (Basic Eavesdropper)} compared with the \textit{Traditional DP Protection} under different privacy budgets $\epsilon$, where the channel SNR is set to 20 dB.}
\label{fig.LPIPS_compareDP}
\end{center}
\vskip -0.3in
\end{figure}

\begin{figure}[t]
\begin{center}
\centerline{\includegraphics[width=0.95\linewidth]{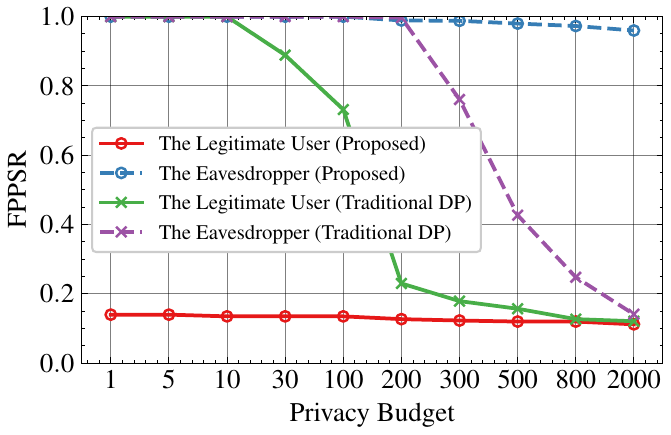}}
\caption{The FPPSR performance of the \textit{Proposed System (Basic Eavesdropper)} compared with the \textit{Traditional DP Protection} under different privacy budgets $\epsilon$, where the channel SNR is set to 20 dB.}
\label{fig.FPPSR_compareDP}
\end{center}
\vskip -0.3in
\end{figure}

\subsection{Comparisons with Traditional DP Protection}


In this subsection, we compare the performance of the \textit{Proposed System (Basic Eavesdropper)} with the \textit{Traditional DP Protection} benchmark \cite{dwork2006differential,li2021differentially} under different privacy budgets $\epsilon$ in an equal-SNR (20 dB) wiretap channel scenario. 

As shown in Figs.~\ref{fig.LPIPS_compareDP} and~\ref{fig.FPPSR_compareDP}, our proposed method consistently outperforms the \textit{Traditional DP Protection} benchmark in terms of both legitimate user performance and privacy protection against the eavesdropper.
Specifically, the \textit{Traditional DP Protection} benchmark introduces genuine DP noise to the private latent codes but suffers from inherent non-invertibility. 
As a result, the legitimate user's reconstruction performance deteriorates significantly, approaching the eavesdropper's performance. 
At low privacy budgets, the legitimate user struggles to reconstruct high-fidelity images. 
Conversely, at high privacy budgets, the eavesdropper can still reconstruct images with high quality.
More concretely, at low privacy budgets, both the legitimate user and the eavesdropper in the \textit{Traditional DP Protection} benchmark show similarly poor LPIPS and FPPSR performance. At high privacy budgets, however, the eavesdropper achieves good LPIPS and FPPSR results, indicating effective image reconstruction and low system security.
This shows that the \textit{Traditional DP Protection} benchmark fails to provide effective privacy protection and high legitimate user performance in wiretap channel scenarios.

In contrast, our proposed method introduces the DP noise with learnable pattern, allowing the legitimate user to effectively mitigate its effect through the DP deprotection module. 
The results demonstrate that our method achieves significant improvements in both system security and task performance compared with the \textit{Traditional DP Protection} benchmark.
%

\section{Summary and Future Work}

This paper proposed a DP-based secure SemCom system over wiretap channels. 
Specifically, an NN-based DP protection module was introduced, where DP noise with learnable pattern was selectively added to private latent codes to provide privacy protection while maintaining high reconstruction quality for the legitimate user. 
To further enhance this process, a discriminator was employed to guide the noise generation toward resembling genuine DP noise, thereby enabling an approximate DP guarantee. 
At the receiver, a corresponding DP deprotection module was designed to effectively mitigate the effect of the introduced noise, enabling reliable image recovery for the legitimate user.
Building on GAN inversion, the proposed fine-grained privacy protection strategy reduced computational overhead and model complexity compared with full-space DP protection.
%
In addition, tunable privacy budgets provided flexible control over the system's security levels, which allowed the proposed system to produce either chaotic or slightly misleading images for the eavesdropper.
Experimental results confirmed that our system outperformed both the previous DP-based method and direct transmission in terms of system security and task performance, making it a practical and robust solution for secure SemCom.
%

This work can be extended in several promising research directions.
First, to more effectively mislead potential eavesdroppers, future research could explore approaches for generating natural-looking protected images even under low privacy budgets.
Second, incorporating robust anti-jamming strategies could help defend against semantic jamming attacks launched by malicious adversaries, thereby mitigating performance degradation for legitimate users.



\bibliographystyle{IEEEtran}

\bibliography{myref}


\vspace{12pt}
\end{CJK}
\end{document}